\documentclass[aps,prd,12pt,nofootinbib]{revtex4}

\usepackage{amsfonts}
\usepackage{amsmath,epsfig,bm}
\usepackage{graphicx}
\usepackage{color}

\usepackage[setpagesize=false]{hyperref}

\usepackage{tabularx}  
\newcolumntype{s}{>{\centering\arraybackslash\hsize=.666\hsize}X} 
\newcolumntype{b}{>{\centering\arraybackslash\hsize=1.333\hsize}X} 

\usepackage{epsfig}

\newcommand{\be}{\begin{equation}}
\newcommand{\ee}{\end{equation}}
\newcommand{\ba}{\begin{eqnarray}}
\newcommand{\ea}{\end{eqnarray}}
\newcommand{\bd}{\begin{displaymath}}
\newcommand{\ed}{\end{displaymath}}

\begin{document}

\title{Baryon Number Fluctuations from a Crossover Equation of State Compared to Heavy-Ion Collision Measurements in the Beam Energy Range $\sqrt{s_{NN}}$ = 7.7 to 200 GeV}

\author{M. Albright,$^1$ J. Kapusta,$^1$ and C. Young$^2$}
\affiliation{$^1$School of Physics \& Astronomy, University of Minnesota, Minneapolis, MN 55455,USA\\
$^2$National Superconducting Cyclotron Laboratory, Michigan State University, East Lansing, MI 48824, USA}
\

\begin{abstract}
Fluctuations of the proton number distribution in central Au-Au collisions have been measured by the STAR collaboration in a beam energy scan at the Relativistic Heavy Ion Collider (RHIC).  The motivation is a search for evidence of a critical point in the equation of state.  It was found that the skewness and kurtosis display an interesting energy dependence.  We compare these measurements to an equation of state which smoothly interpolates between an excluded volume hadron resonance gas at low energy density to a perturbative plasma of quarks and gluons at high energy density.  This crossover equation of state agrees very well with the lattice QCD equation of state.  The crossover equation of state can reproduce the data if the fluctuations are frozen at a temperature significantly lower than the average chemical freeze-out.     
\end{abstract}

\maketitle

\parindent=20pt


\section{Introduction}
\label{Intro}

The motivation for the beam energy scan at the Relativistic Heavy Ion Collider (RHIC) is to search for a critical end point in the QCD phase diagram.  Various models and theoretical arguments suggest that there is a curve in the plane of temperature $T$ versus baryon chemical potential $\mu$ representing a line of first-order phase transitions.  This curve terminates in a second-order phase transition at some $T_c$ and $\mu_c$.  There is no agreement on the numerical values, but the expectation is that $T_c$ is less than 160 MeV and $\mu_c$ is greater than a few hundred MeV.  (For reviews see \cite{Stephanov} and \cite{MohantyQM}.)  More generally one would like to create matter in heavy-ion collisions with moderate temperatures and high baryon densities to study the type of matter that exists in proto-neutron and neutron stars.  The challenge is entropy.  One needs high enough collision energies to create matter at high energy density, but the collision energies cannot be too high because the entropy per baryon also increases with collision energy and that means high temperatures and low chemical potentials.

Recently, the STAR collaboration at RHIC has reported measurements of the moments of net-proton (proton minus anti-proton) multiplicity distributions in Au-Au collisions during the first beam energy scan \cite{STAR_BES}.  These moments have been proposed as good observables for critical behavior \cite{3rdmoments, 4thmoments, nongaussian}.  The measurements were performed at beam energies $\sqrt{s_{NN}}$ = 7.7, 11.5, 19.6, 27, 39, 62.4, and 200 GeV per nucleon pair.  This is remarkable as the two lowest energies are below the injection energy of 19.6 GeV.  Here we focus on the most central collisions which are in the range 0-5\%.

In a previous paper \cite{matchingpaper}, we reported on the construction of several equations of state which smoothly interpolate between a hadron resonance gas with excluded volumes to a perturbative QCD plasma of quarks and gluons as the temperature and/or chemical potential is raised.  The handful of parameters were adjusted to reproduce the pressure and interaction measure (also called trace anomaly) calculated with lattice QCD for $T$ between 100 and 1000 MeV with $\mu$ = 0.  With no further free parameters the crossover equation(s) of state represented the lattice results at $\mu$ = 400 MeV just as well.  In this paper we compare the third and fourth moments of the baryon distribution from these crossover equations of state to the STAR data.  Fukushima \cite{Fukushima} made a similar comparison using a point particle hadron resonance gas and a relativistic mean field model with vector and scalar interactions.  Bors{\'a}nyi {\it et al.} \cite{Borsanyi2014PRL} compared their lattice QCD results to the second moments at $\sqrt{s_{NN}}$ = 27, 39, 62.4, and 200 GeV and to the third moment obtained by averaging over these energies.

The outline of this paper is as follows.  In section \ref{soundspeed} we briefly review the equations of state used here.  We also compare them to each other and to lattice QCD calculations of the sound speed as a function of temperature, both for $\mu$ = 0 and $\mu$ = 400 MeV.   We make similar comparisons for the fourth moment, or kurtosis, for $\mu$ = 0 in section \ref{kurtosis}.  In section \ref{chemical} we make comparisons to the STAR data assuming that the fluctuations are determined at the time of average chemical freeze-out.  In order to represent the STAR data better, in section \ref{postchemical} we consider the possibility that the fluctuations are determined significantly after the average chemical freeze-out.  Our conclusions are  contained in section \ref{conclusion}.

\section{Speed of Sound}
\label{soundspeed}

In \cite{matchingpaper} we constructed a pressure $P(T,\mu)$ which includes a hadronic piece $P_h$, a perturbative QCD piece $P_{qg}$, and a switching function $S$.
\be
P(T,\mu) = S(T,\mu) P_{qg}(T,\mu)  + \left[1 - S(T,\mu)  \right] P_h(T,\mu)
\label{switch}
\ee
The hadronic piece consists of a resonance gas comprising all the known hadrons as presented in the Particle Data Tables \cite{pdg2012}; an explicit list may be found in the appendix of \cite{matchingpaper}.  In one case the hadrons were treated as point particles.  We also used two excluded volume models for the resonance gas. For model I, the assumption is that the volume excluded by a hadron is proportional to its energy $E$ with the constant of proportionality $\epsilon_0$ (dimension of energy per unit volume) being the same for all species.  
For model II, the assumption is that the volume excluded by a hadron is proportional to its mass $m$.  It is assumed in both models that hadrons are deformable so that there is no limitation by a packing factor as there would be for rigid spheres, for example.  We refer to these three hadronic equations of state as pt, exI and exII. Philosophically these excluded volume models are similar but the mathematics to compute their partition functions is rather different.  Quantum statistics are used for the hadronic piece of the equation of state.

For the perturbative QCD piece we use the latest calculation which is valid up order $\alpha_s^3 \ln \alpha_s$.  An explicit expression may be found in the appendix of \cite{matchingpaper}.  This piece involves two parameters, both relating to the energy scale used in the renormalization group running coupling.

The $S(T,\mu)$ is a switching function which must approach zero at low temperatures and chemical potentials and approach one at high temperatures and chemical potentials.  The switching function must also be infinitely differentiable to avoid introducing first, second, or higher-order phase transitions.  The following functional form was chosen
\ba
 S(T, \mu) &=& \exp\{-\theta(T, \, \mu)\} \nonumber \\  
\theta(T, \mu) &=& \left[ \left( \frac{T}{T_0}\right)^r +   \left(\frac{\mu}{3 \pi T_0} \right)^r  \right]^{-1} 
\ea
with integer $r$.   This function goes to zero faster than any power of $T$ as $T \rightarrow 0$ (when $\mu = 0$).  It has two parameters, $T_0$ and $r$. 

We now have two parameters in the switching function, two parameters in the perturbative QCD equation of state, and one parameter in the excluded volume equation of state ($\epsilon_0$ not necessarily the same for both models).  We did a search on the parameters in each of the three models to obtain the best overall chi-square fit to both the pressure $P/T^4$ and the trace anomaly $(\epsilon - 3P)/T^4$ to the data in ref. \cite{Borsanyi2010JHEP}.  The point particle model best fits resulted in $r=4(5)$ and $T_0=145.3(157.4)$ MeV with a chi-squared per degree of freedom of 0.56(0.62).  The exI model best fit resulted in $r=5$, $T_0=177.1$ MeV, and $\epsilon_0^{1/4} = 306.5$ MeV with a chi-squared per degree of freedom of 0.34.  Nearly identical to that, the exII model best fit resulted in $r=5$, $T_0=177.6$ MeV, and $\epsilon_0^{1/4} = 279.7$ MeV with a chi-squared per degree of freedom of 0.34.  The value of $T_0$ for the point hadron gas is significantly smaller than for the excluded volume models, and so the switching from hadrons to quarks and gluons occurs at a lower temperature.  The reason is that $P_h$ for the point hadron model grows much faster with $T$ than for the excluded volume models, and that fast growth must be cut-off by the switching function.  An unnatural consequence is that there is a minor dip in the trace anomaly near a temperature of 115 MeV in the point particle model.  Furthermore, if an exponential mass spectrum of the Hagedorn type was used instead of only the known hadrons as presented in the Particle Data Tables, the $P_h$ for point hadrons would not even be defined above the Hagedorn temperature $T_H \approx 160$ MeV.  With the inclusion of excluded volumes for the hadrons this is not the case.   

An interesting physical result of the excluded volume model fits is the resulting hard core radius of the proton and neutron.  For exI it is 0.580 fm while for exII it is 0.655 fm, very sensible numbers.

Although we show some results with the point hadron resonance gas in what follows, it does have a fatal flaw when used as the hadronic piece in a crossover equation of state.  If the point particle hadron resonance gas has a mass spectrum which grows exponentially {\it a la} Hagedorn, rather than with a large but finite number of hadronic resonances, then the hadronic piece has a pressure which either diverges or reaches a maximum at the Hagedorn temperature $T_H \approx 160$ MeV.  Then the crossover or switching method of Eq. (\ref{switch}) fails to produce a smooth crossover. 

\begin{figure}[thp]
\begin{center}
\includegraphics[width=0.8\linewidth]{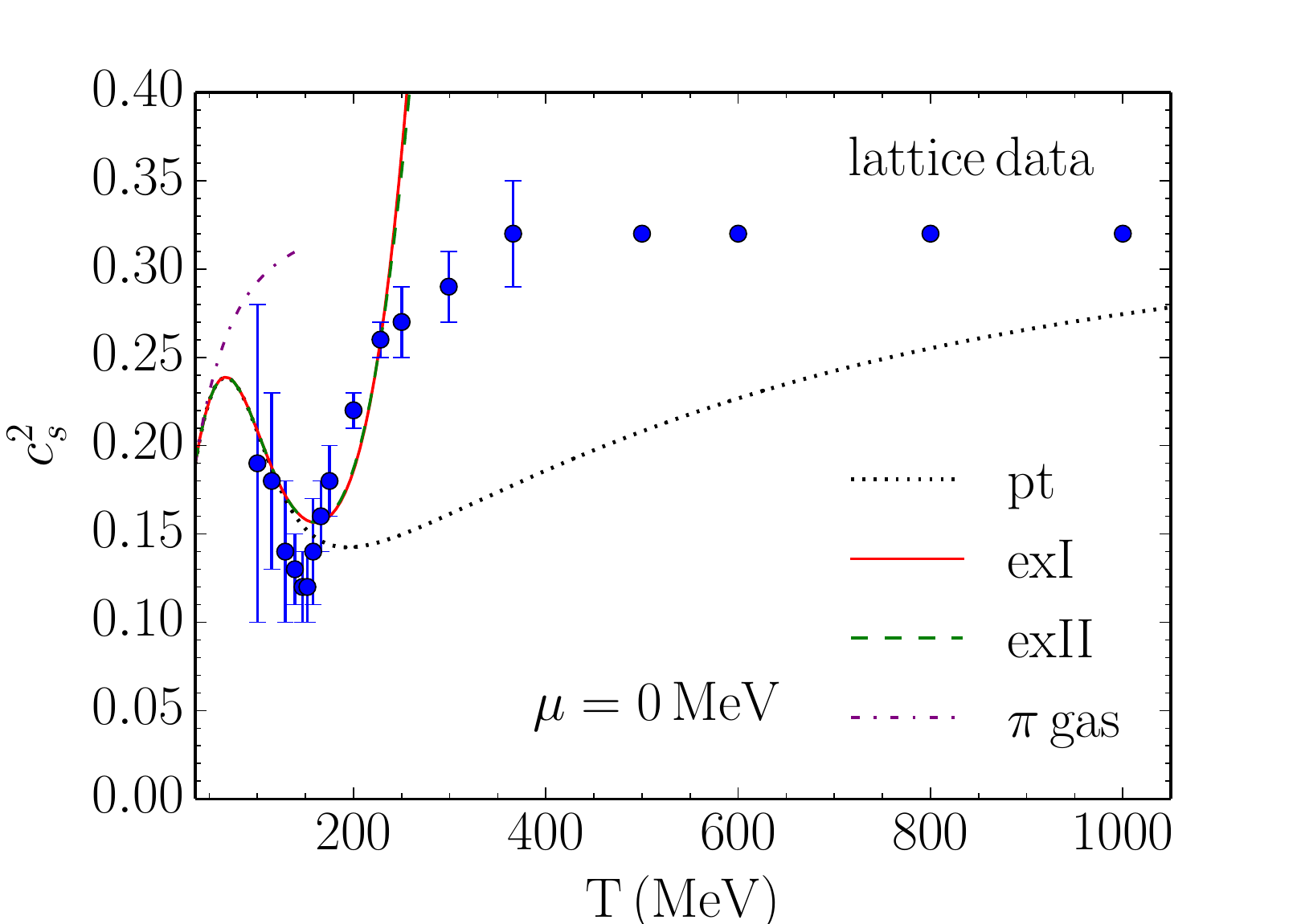}
\includegraphics[width=0.8\linewidth]{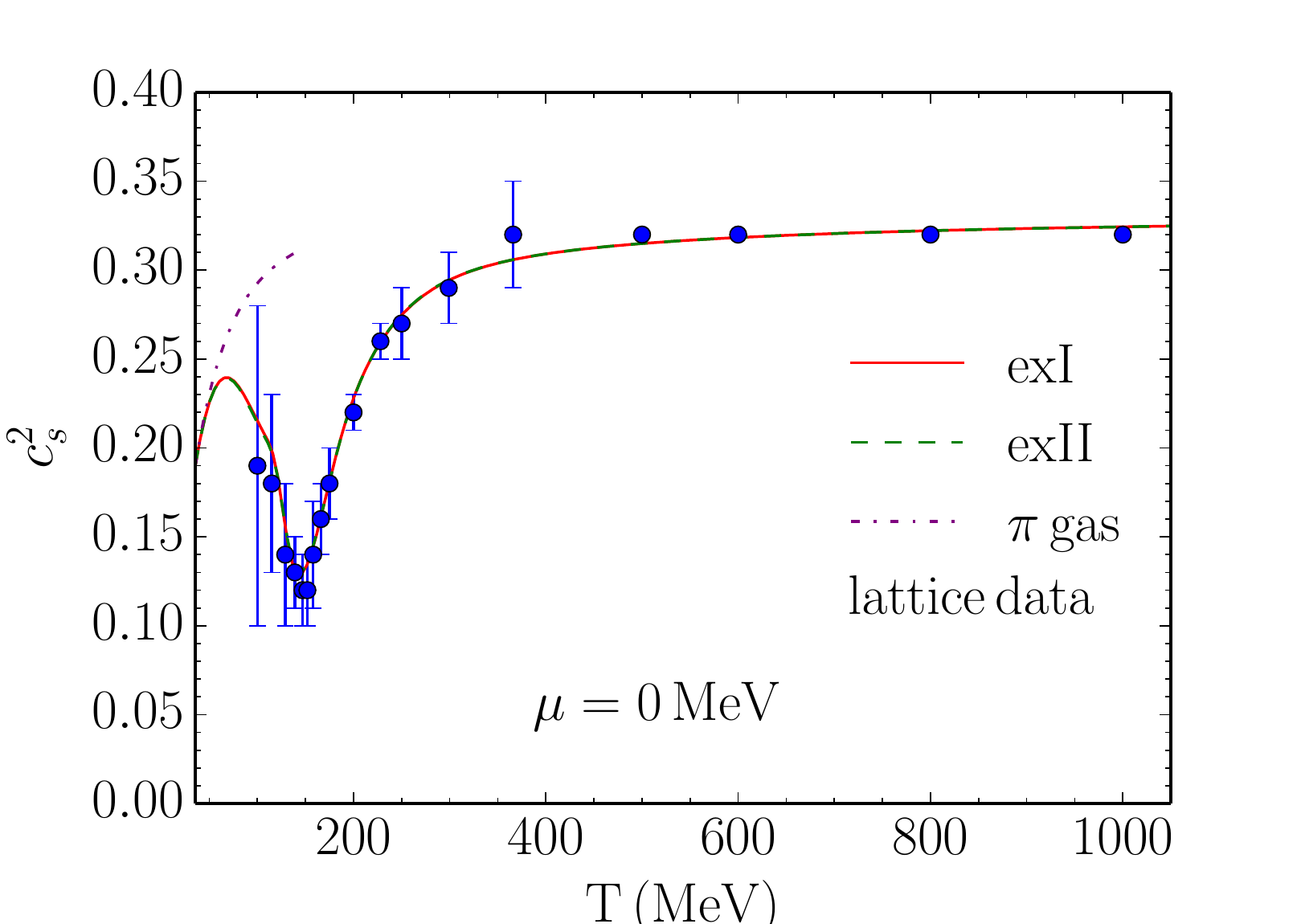}
\caption{(color online) The square of the sound speed as a function of temperature for zero baryon chemical potential.  The points are from lattice QCD \cite{Borsanyi2010JHEP}.  The dashed line is for a noninteracting massive pion gas. The top panel shows the sound speed for hadronic resonance gases, including those for point hadrons and for the two excluded volume models.  The bottom panel shows the full crossover equations of state.}
\end{center}
\label{crossover_cs2_0}
\end{figure}

\begin{figure}[thp]
\begin{center}
\includegraphics[width=0.8\linewidth]{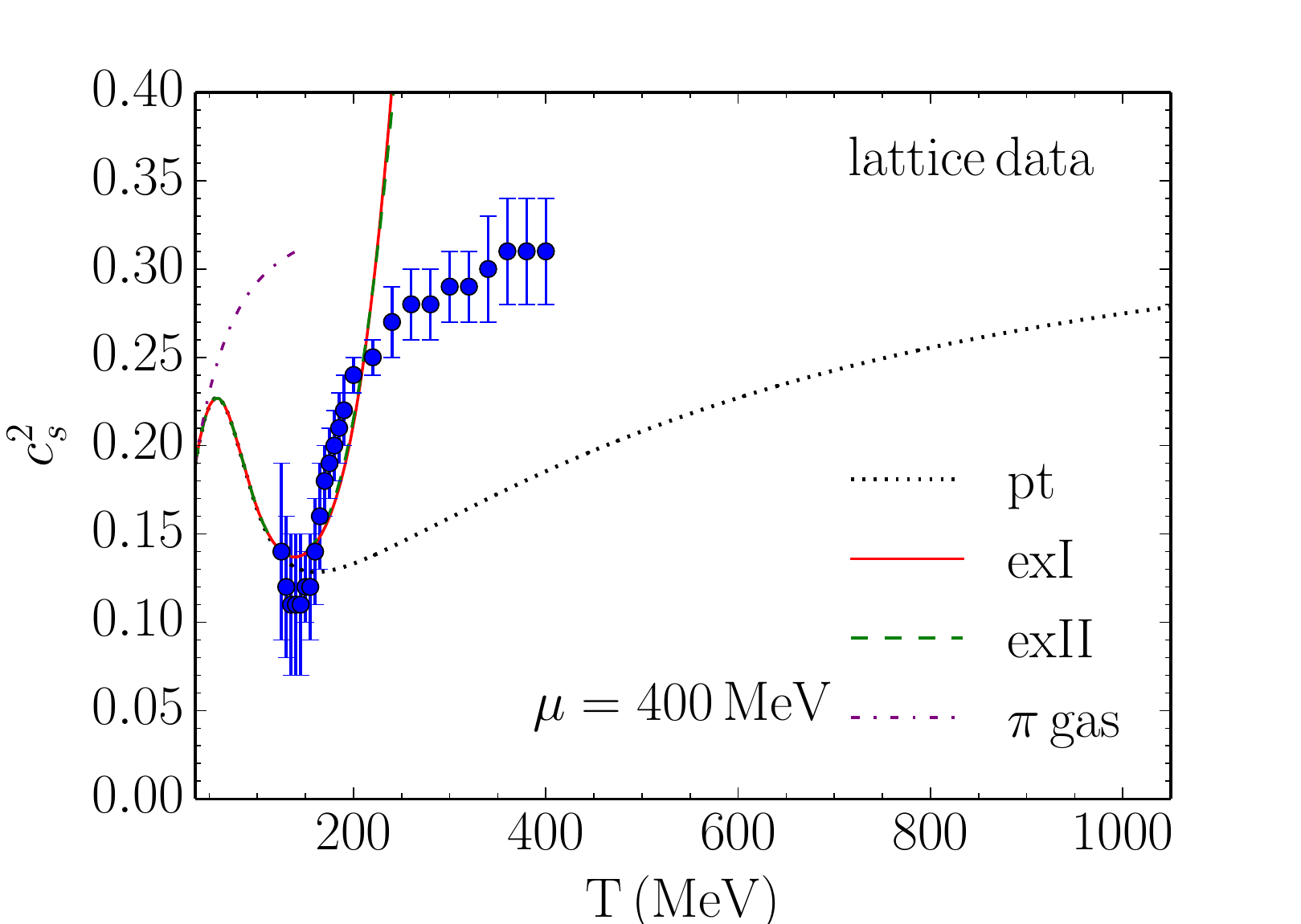}
\includegraphics[width=0.8\linewidth]{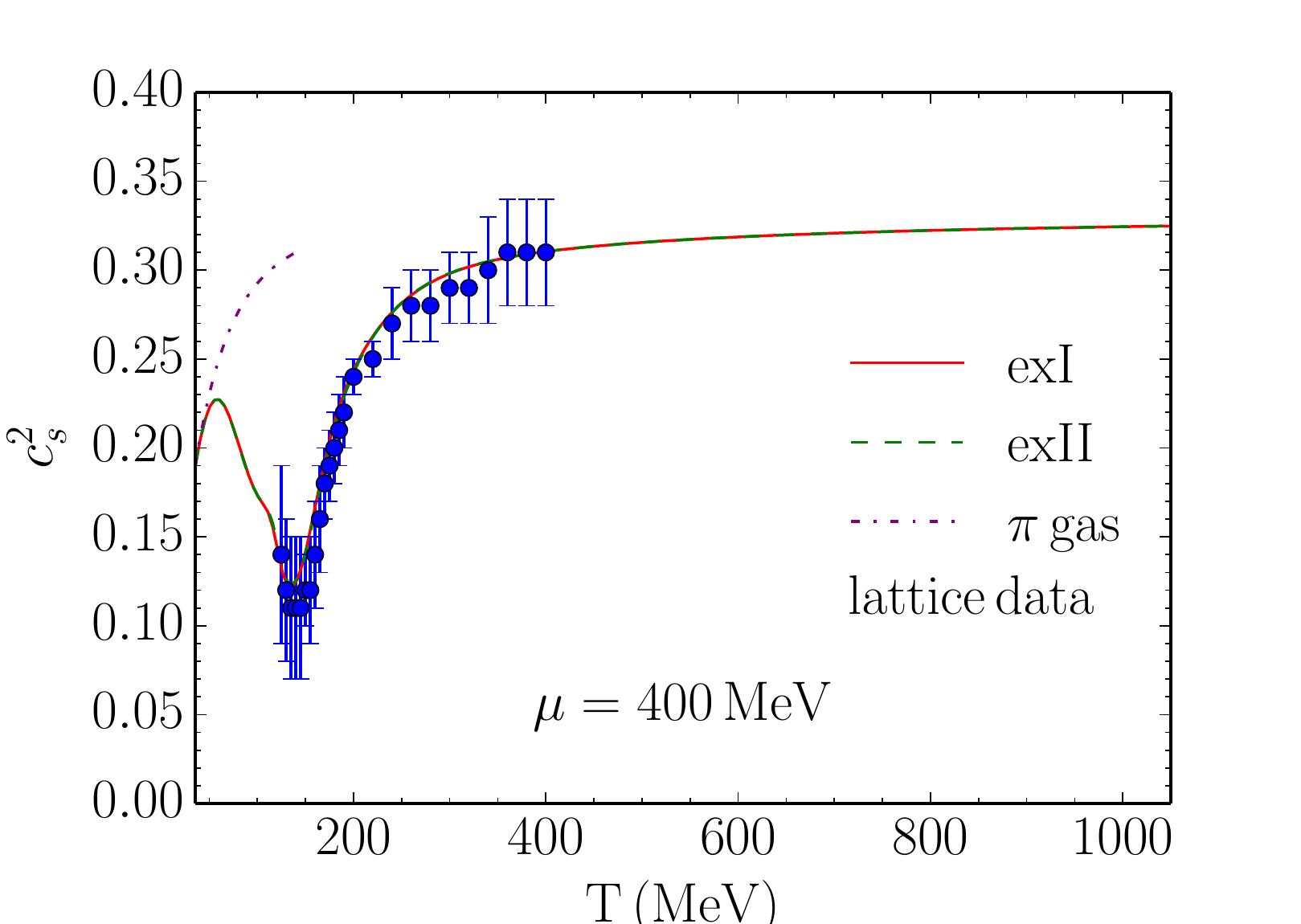}
\caption{(color online) The square of the sound speed as a function of temperature for a baryon chemical potential of 400 MeV.  The points are from lattice QCD \cite{Borsanyi2012}.  The dashed line is for a noninteracting massive pion gas. The top panel shows the sound speed for hadronic resonance gases, including those for point hadrons and for the two excluded volume models.  The bottom panel shows the full crossover equations of state.}
\end{center}
\label{crossover_cs2_400}
\end{figure}

A more sensitive test of the agreement between the crossover models and lattice QCD is the speed of sound given by $c_s^2 = \partial P/\partial \epsilon$, where the derivative is taken at constant entropy per baryon.  In terms of the susceptibilities
\be
\chi_{xy} = \frac{\partial^2 P(T,\mu)}{\partial x \partial y}
\ee
where $x$ and $y$ may be $T$ or $\mu$, it is
\be
c_s^2 = \frac{n_B^2 \chi_{TT} - 2 s n_B \chi_{\mu T} + s^2 \chi_{\mu\mu}}{w (\chi_{TT} \chi_{\mu\mu} - \chi^2_{\mu T})} \,.
\label{speed}
\ee
Whereas the trace anomaly depends on first-order derivatives of the pressure, the sound speed depends also on second-order derivatives.  A comparison to lattice QCD is shown in Fig. 1 for $\mu=0$ \cite{Borsanyi2010JHEP} and in Fig. 2 for $\mu=400$ MeV \cite{Borsanyi2012}.  In both figures the top panel shows the hadronic piece only, in other words set $S(T,\mu)=0$.  (It might be mentioned that the excluded volume parameter $\epsilon_0$ was adjusted to try to give a better fit to the pressure and trace anomaly  of lattice QCD.)  The bottom panel shows the full crossover equation of state.  The crossover equation of state using point hadrons has an unnatural temperature dependence around 115 MeV (see comment above) and so we do not show it in these figures.  In both panels the dashed curve shows the sound speed for a massive non-interacting pion gas, which is the natural limit as the temperature goes to zero.  The hadronic piece only gives a qualitative representation of the lattice data, where the full crossover equation of state reproduces it very well.  

It is possible to switch smoothly between the excluded volume and the perturbative QCD equations of state thanks to the existence of a region of overlap where both equations of state match the lattice data. Ultimately, the switching is possible not because of the choice of $S(T, \mu)$, but by the improved description of the physics of hadronic gas making it accurate to higher temperatures. 

\section{Skewness and Kurtosis}
\label{kurtosis}

Let $Q$ represent some conserved charge.  In an ensemble of events it will have an average value $\bar{Q}$ and a variance $\sigma^2$.   The next higher moments are the skewness
\be
S=\frac{\langle \delta Q \rangle^3}{\sigma^3}
\ee
and kurtosis
\be
\kappa=\frac{\langle \delta Q \rangle^4}{\sigma^4}-3
\ee
where $\delta Q = Q - \bar{Q}$.  Here we focus on the net baryon number.  For a finite size system, such as the matter formed in heavy-ion collisions, it is convenient and useful to consider the scaled skewness and kurtosis which are intensive quantities.  For a system in equilibrium they are
\be
S \sigma = T \frac{\chi_{\mu\mu\mu}}{\chi_{\mu\mu}}
\ee
and
\be
\kappa \sigma^2 = T^2 \frac{\chi_{\mu\mu\mu\mu}}{\chi_{\mu\mu}} \,.
\ee
One general statement that can be made is $S \sigma = 0$ if the chemical potential vanishes.  The reason is that the pressure is an even function of the chemical potential, hence all odd derivatives vanish when $\mu = 0$.

Consider a point particle hadron resonance gas.  When Boltzmann statistics are adequate for the baryons, the pressure takes the form
\be
P(T,\mu) = P_b(T) \cosh(\mu/T) + P_m(T)
\ee
in an obvious notation.  Then $S \sigma = \tanh(\mu/T)$ and $\kappa \sigma^2 = 1$.  This is a very strong prediction.  For comparison, consider a noninteracting gas of massless quarks and gluons.  The pressure takes the form
\be
P(T,\mu) = a_0 T^4 + a_2 T^2 \mu^2 + a_4 \mu^4
\ee
where $a_0$, $a_2$, and $a_4$ are constants.  Then $S \sigma = 12 a_4 \mu T / (a_2 T^2 + 6 a_4 \mu^2)$ and $\kappa \sigma^2 = 12 a_4 T^2 / (a_2 T^2 + 6 a_4 \mu^2)$.  The dependence on the ratio $\mu/T$ is very different for these two equations of state.

\begin{figure}[thp]
\begin{center}
\includegraphics[width=0.8\linewidth]{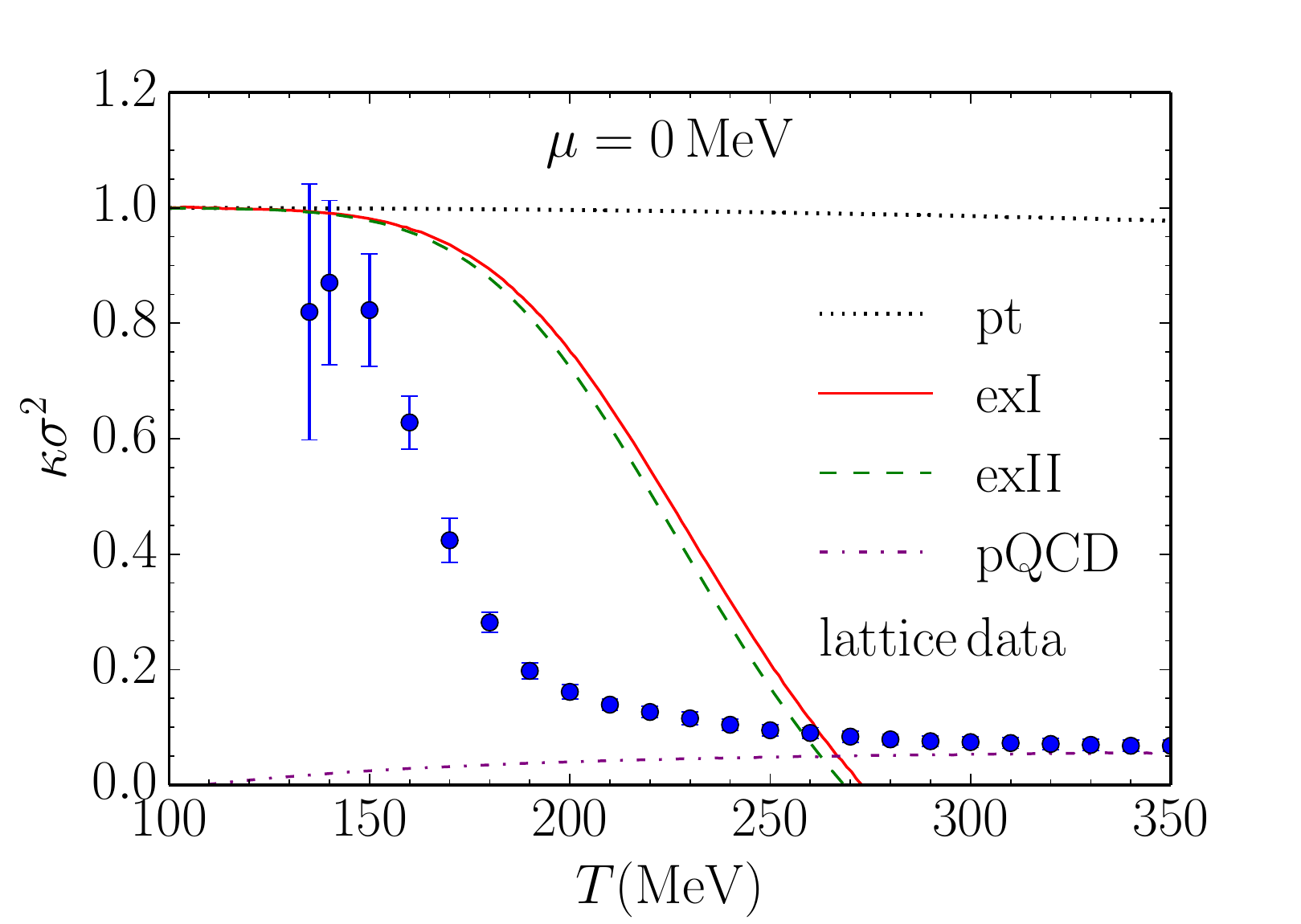}
\includegraphics[width=0.8\linewidth]{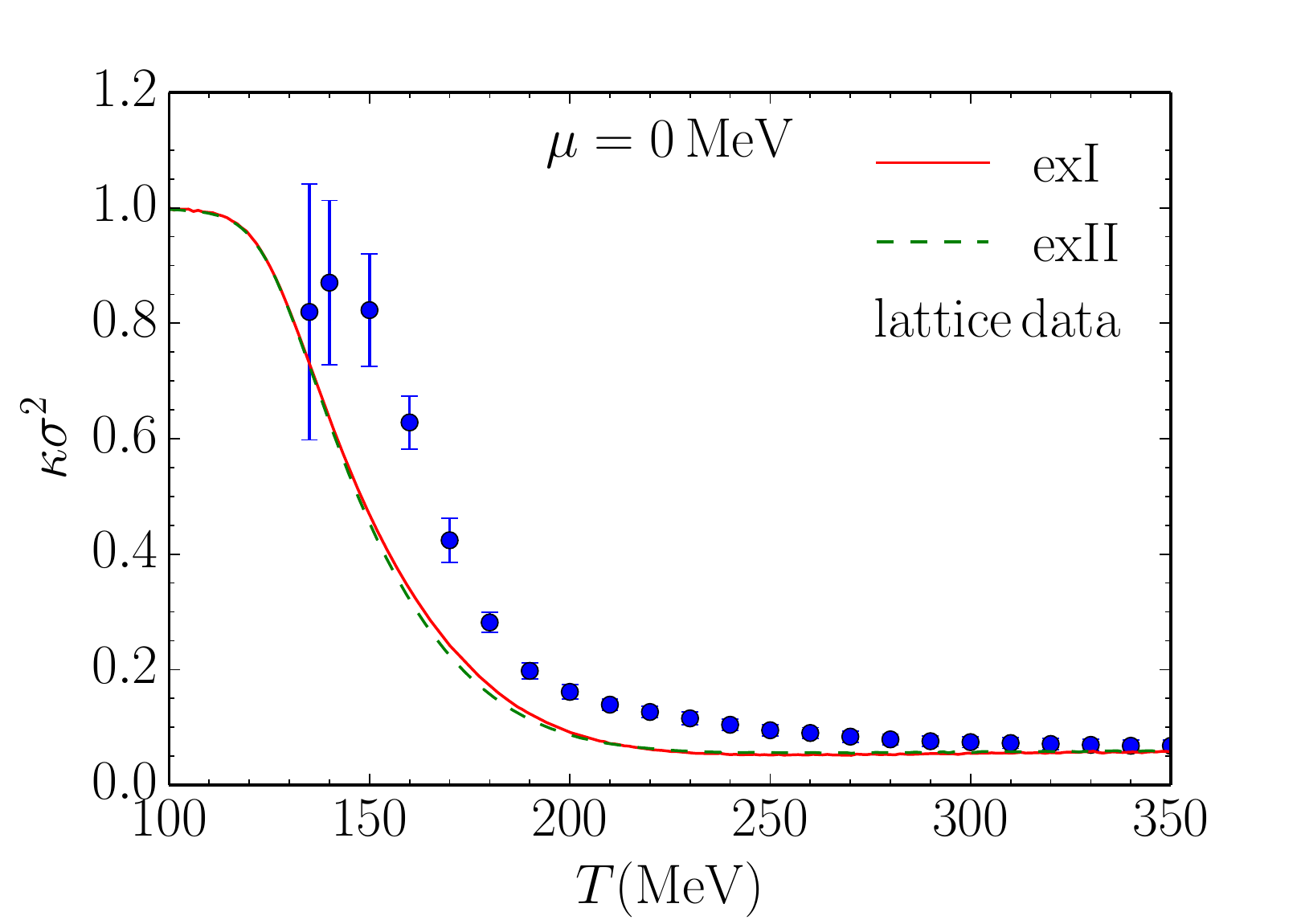}
\caption{(color online) The kurtosis as a function of temperature for zero baryon chemical potential.  The points are from lattice QCD \cite{Borsanyi2013PRL}.  The top panel shows the kurtosis for hadronic resonance gases, including those for point hadrons and for the two excluded volume models.  It also shows the purely perturbative QCD result for quarks and gluons.  The bottom panel shows the full crossover equations of state.}
\end{center}
\label{crossover_kurtosis_0}
\end{figure}

The kurtosis has been computed by lattice QCD \cite{Borsanyi2013PRL} and the results are shown in Fig. 3 (the skewness is obviously zero).  The top panel shows a comparison to the point hadron resonance gas and the two excluded volume hadron resonance gases.  The kurtosis for the point particle hadron resonance gas is essentially 1, but deviates very slightly with increasing temperature due to quantum statistics.  The kurtosis for the excluded hadron resonance gas does decrease significantly with temperature, but it does not quantitatively reflect the lattice QCD kurtosis.  Also shown in the top panel is the kurtosis from perturbative QCD.  Deviation from a noninteracting gas of massless quarks and gluons is evident and is due to the running of the renormalization group coupling $\alpha_s$ with $T$ and $\mu$.  Lattice QCD clearly shows the transition from a hadron resonance gas at low temperature to a plasma of quarks and gluons at high temperature.  The bottom panel shows the comparison with the full crossover equations of state exI and exII.  Qualitatively the agreement is very good, the crossover equations of state having the correct limits when $T$ becomes small and when it becomes large.  However, in the intermediate region between 150 and 
300 MeV, the kurtosis from the crossover equations of state is too small compared to lattice QCD.  This is due to the perturbative equation of state lying below the lattice data. While this deserves further exploration, it is most likely inconsequential for the following sections, where lower temperatures are relevant.

\begin{figure}[tbp]
\begin{center}
\includegraphics[width=0.8\linewidth]{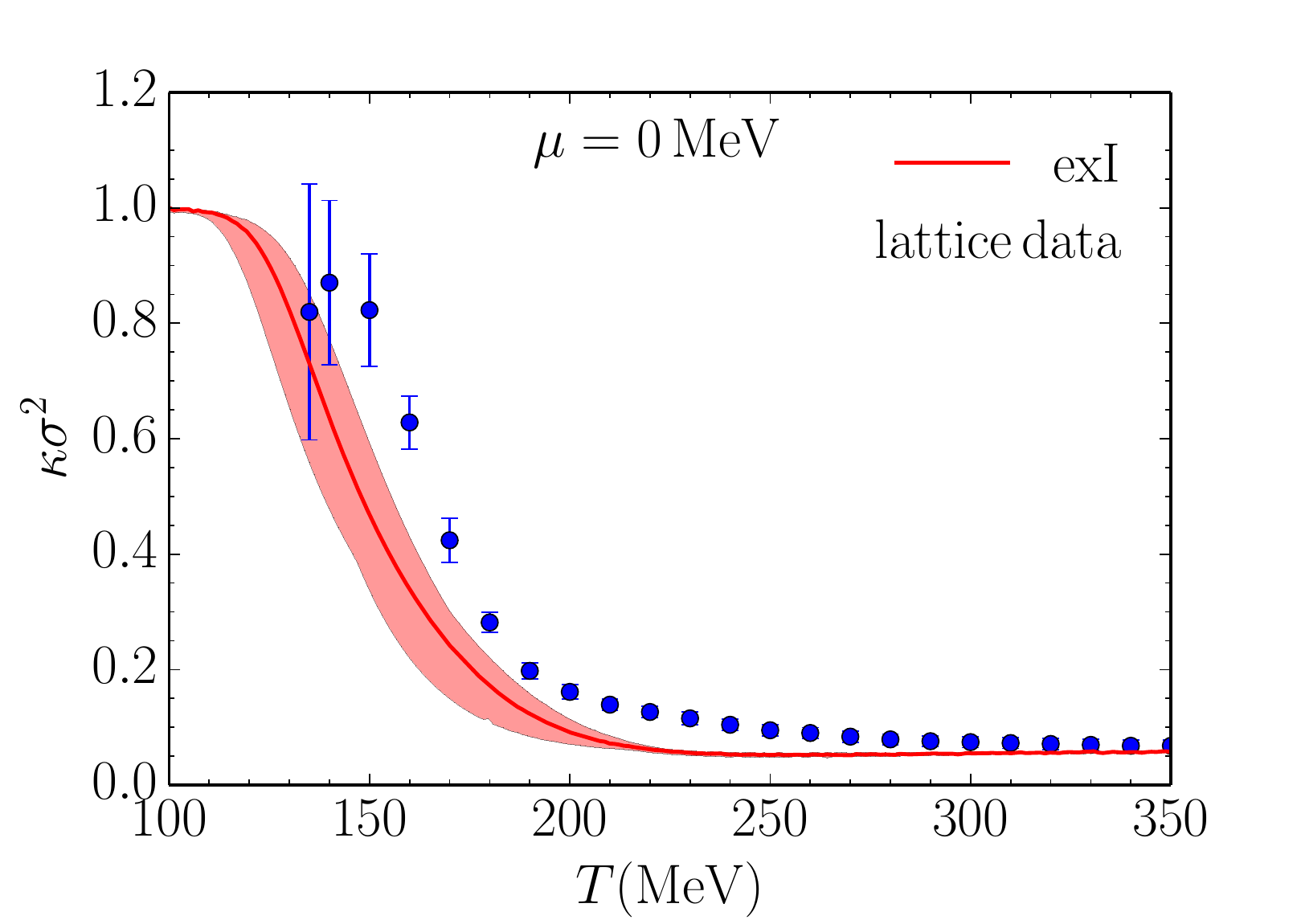}
\caption{(color online) The kurtosis as a function of temperature for zero baryon chemical potential.  The points are from lattice QCD.  The curve is the same as the crossover from the previous figure.   The shaded region shows the uncertainty when fitting the crossover equation of state parameters to lattice QCD at zero chemical potential.}
\end{center}
\label{crossover_kurtosis_0_errorband}
\end{figure}

To see whether the discrepancy can be resolved by the fitting procedure to the pressure and trace anomaly, we show the band of uncertainty in the crossover parameters in Fig. 4.  Clearly it cannot.  Since the kurtosis involves the fourth derivative of the pressure with respect to chemical potential, it is a very sensitive measure of the equation of state.  The discrepancy might be due to a slightly inaccurate parameterization of the crossover between hadrons and quarks and gluons, although the agreement with the pressure, trace anomaly, and sound speed at $\mu =0$ and $\mu = 400$ MeV makes this problematic.

\section{Comparison to STAR Data: Chemical Freeze-Out}
\label{chemical}

In this section we compare to the data taken during the first Beam Energy Scan at RHIC by the STAR collaboration \cite{STAR_BES}.  In order to make the comparison we need to have an estimate of the temperature and chemical potential at the time the fluctuations are frozen out.  Following Fukushima \cite{Fukushima} we use the conditions at the time of average chemical freeze-out as presented in \cite{freezeoutcurve}.  The chemical potential is parameterized as a function of $\sqrt{s_{NN}}$ by
\be
\mu(\sqrt{s_{NN}}) = \frac{d}{1 + e \sqrt{s_{NN}}}
\ee
and then the temperature by
\be
T(\mu) = a - b \mu^2 - c \mu^4
\label{Tchem}
\ee
The five constants in these parameterizations are $a = 0.166$ GeV,  $b = 0.139$ GeV$^{-1}$, $c = 0.053$ GeV$^{-3}$, $d = 1.308$ GeV, and $e = 0.273$ GeV$^{-1}$.

\begin{figure}[thp]
\begin{center}
\includegraphics[width=0.8\linewidth]{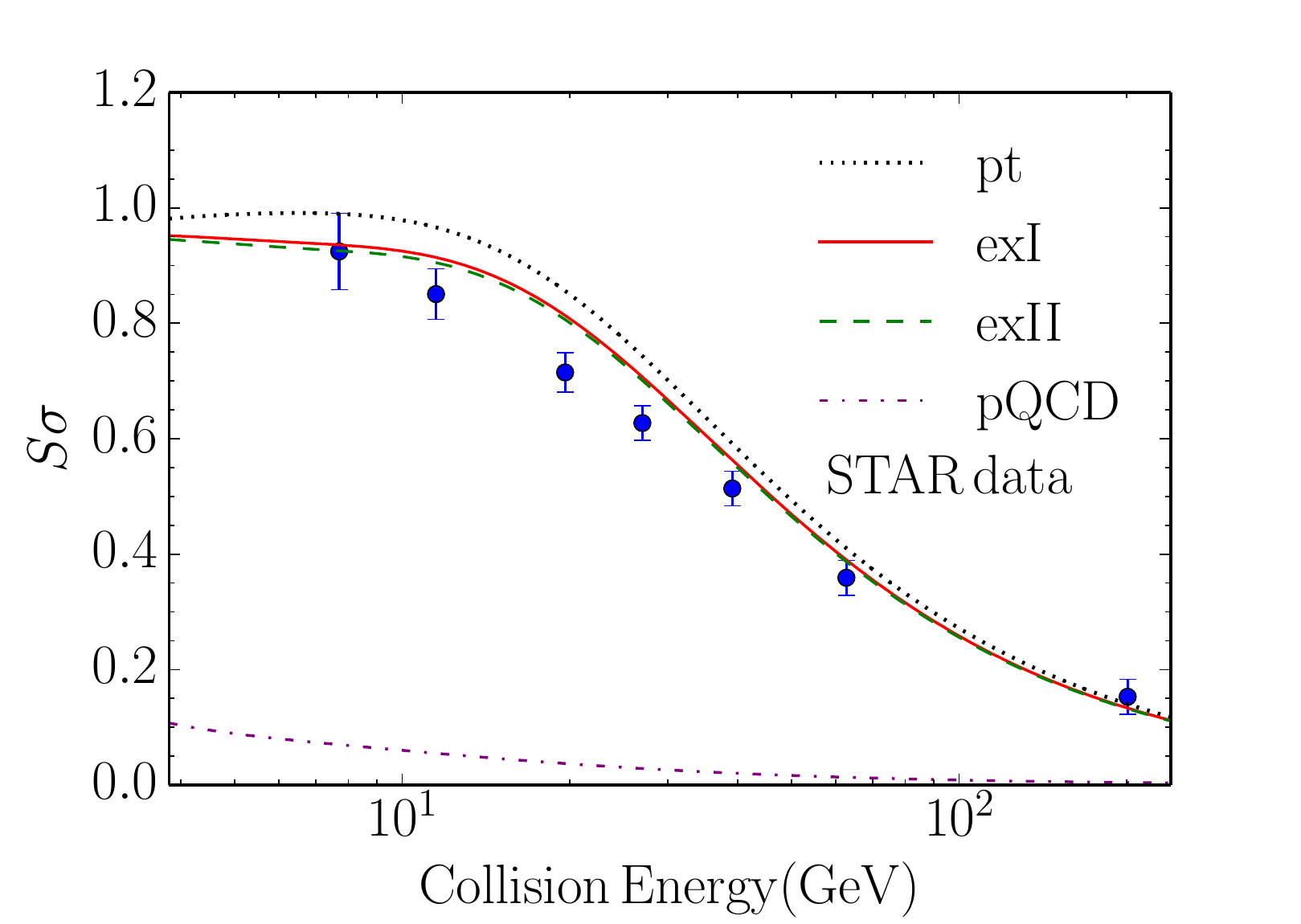}
\includegraphics[width=0.8\linewidth]{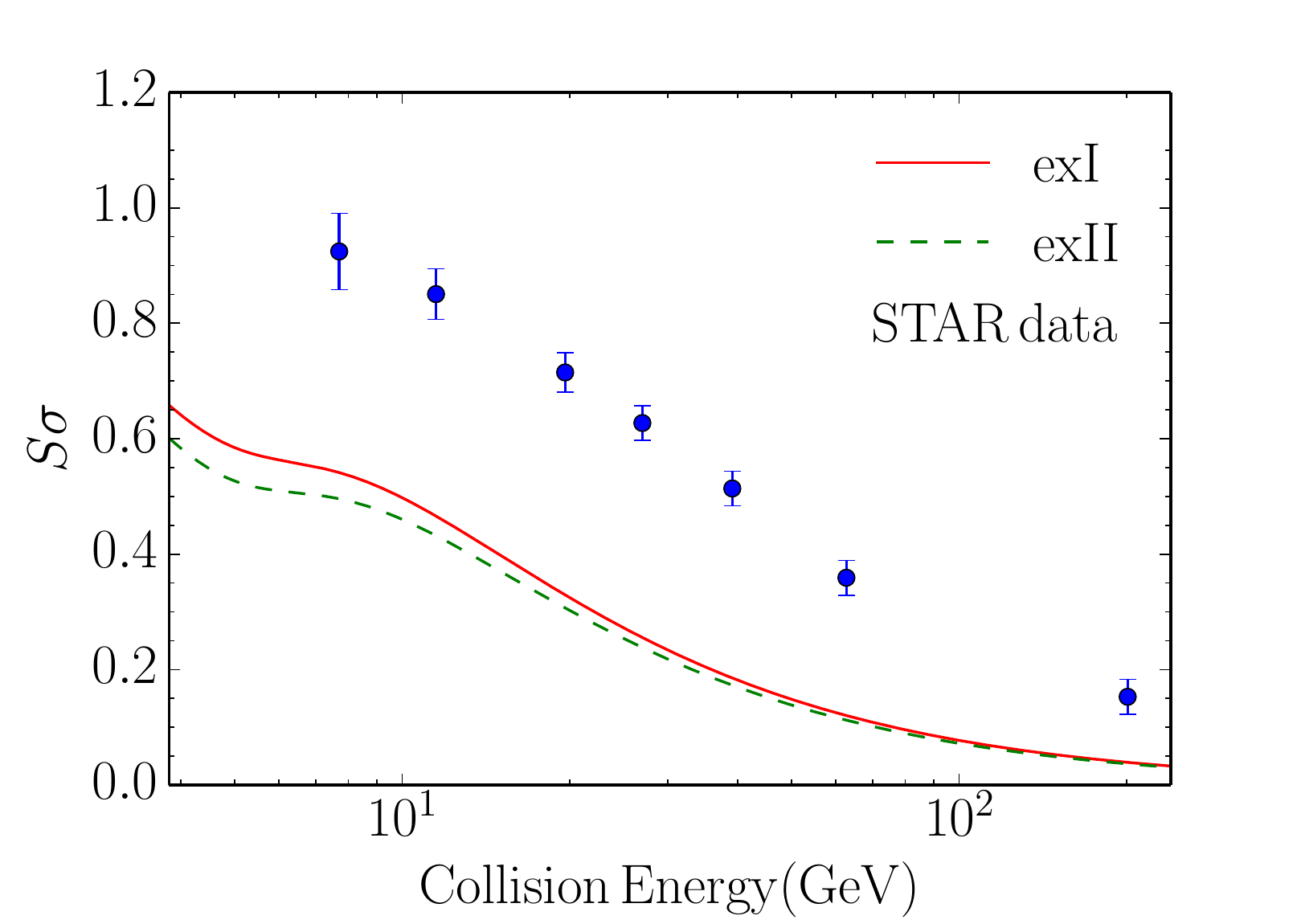}
\caption{(color online) The skewness as a function of collision energy per nucleon pair.  The points are from STAR measurements.  The top panel shows the skewness for excluded volume hadronic resonance gases.  It also shows the purely perturbative QCD result for quarks and gluons.  The bottom panel shows the full crossover equations of state.  The energy dependence of the temperature and chemical potential are determined by the conditions at average chemical freeze-out as in Eq.~(\ref{Tchem}).}
\end{center}
\label{crossover_skewness_STAR_166}
\end{figure}

\begin{figure}[thp]
\begin{center}
\includegraphics[width=0.8\linewidth]{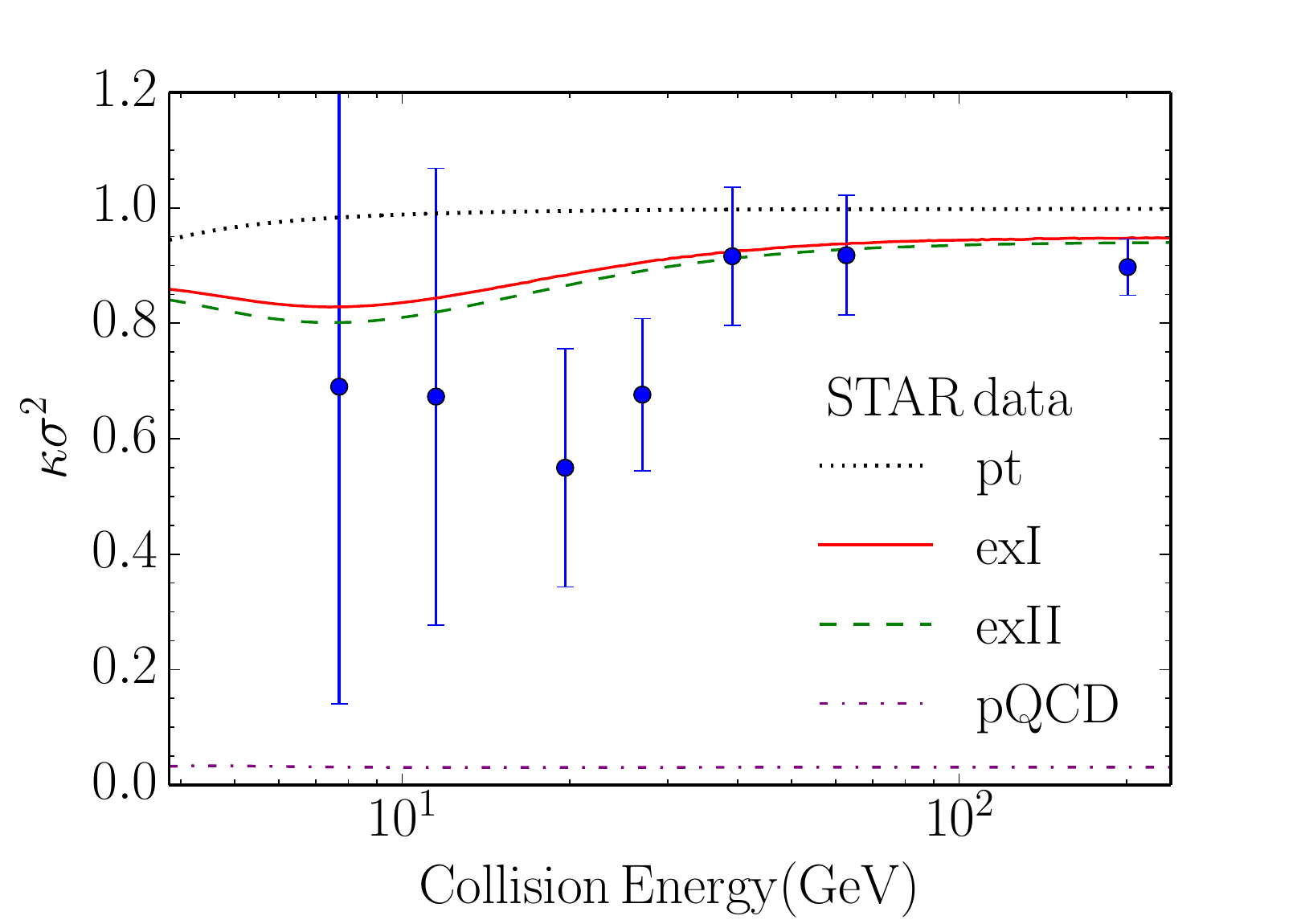}
\includegraphics[width=0.8\linewidth]{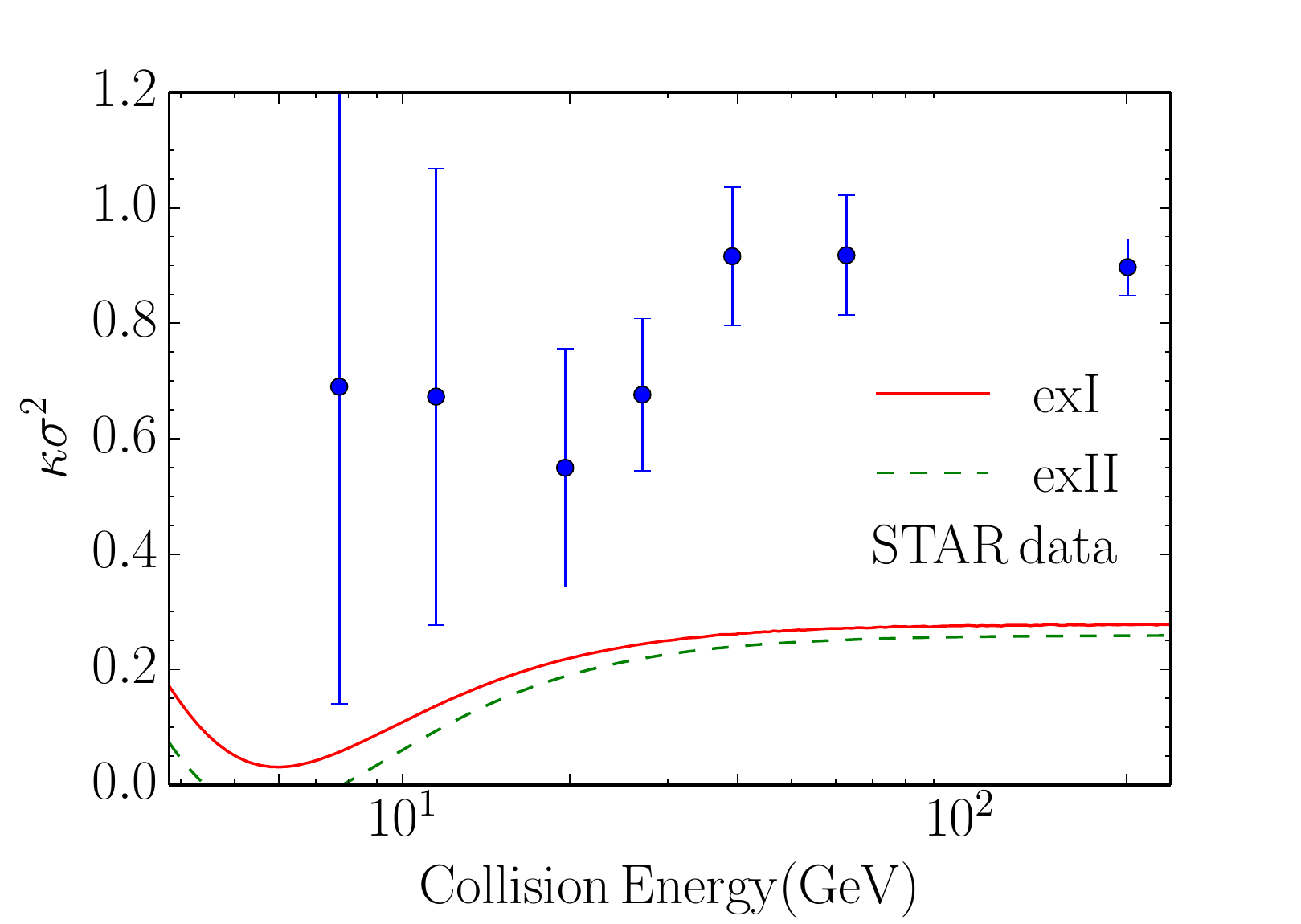}
\caption{(color online) The kurtosis as a function of collision energy per nucleon pair.  The points are from STAR measurements.  The top panel shows the kurtosis for excluded volume hadronic resonance gases.  It also shows the purely perturbative QCD result for quarks and gluons.  The bottom panel shows the full crossover equations of state.  The energy dependence of the temperature and chemical potential are determined by the conditions at average chemical freeze-out as in Eq.~(\ref{Tchem}).}
\end{center}
\label{crossover_kurtosis_STAR_166}
\end{figure}

In Fig. 5 we show the skewness as a function of beam energy.  First, consider the top panel which displays the hadronic and perturbative QCD pieces only.  The perturbative QCD piece is always small and goes to zero with increasing energy.  The reason is that the chemical potential goes to zero with increasing energy, and the skewness vanishes for zero chemical potential as discussed in the previous section.  The point particle hadron gas closely follows the formula $S \sigma = \tanh(\mu/T)$ with only a minor correction for quantum statistics.  Qualitatively it follows the trend of the data.  The two excluded volume models are similar but lie somewhat closer to the data.  The bottom panel shows the skewness from the full crossover equations of state with the excluded volume approximations in the hadronic piece.  (We don't show the crossover with point hadrons for the same reason as with the sound speed.)  These lie significantly below the data and do far worse than the pure hadronic pieces only.  Mathematically the reason for this is that the switching function is reducing the hadronic contribution and increasing the perturbative QCD contribution.

In Fig. 6 we show the kurtosis as a function of beam energy.  The top and bottom panels display the same equations of state as in Fig. 5.  The kurtosis for the point hadron resonance gas would be exactly 1 if classical statistics were used; it drops slightly below 1 at low energy due to the slight effect of quantum statistics.  It cannot be said to represent the data very well, certainly it does not have the dip that the data has at the 19.6 GeV beam energy.  The pure hadronic excluded volume equations of state lie closer to the data but do not reproduce such a strong dip at the aforementioned energy.  

As in the previous section, we show the band of uncertainty in the crossover parameters for exI in Fig. 7 for the skewness (top panel) and kurtosis (bottom panel). These uncertainties are not large enough to explain the STAR data.  

This leaves us with a puzzle.  The crossover equations of state exI and exII do an excellent job of reproducing the pressure, the trace anomaly, and the sound speed as calculated with lattice QCD as a function of temperature between 100 and 1000 MeV and for a baryon chemical potential of 0 and 400 MeV.  They also reproduce the lattice results relatively well for the kurtosis for zero chemical potential.  Yet they do not adequately describe the STAR data as well as the purely hadronic equation of state, which does a much worse job of reproducing the lattice results.  Why is that?

\begin{figure}[thp]
\begin{center}
\includegraphics[width=0.8\linewidth]{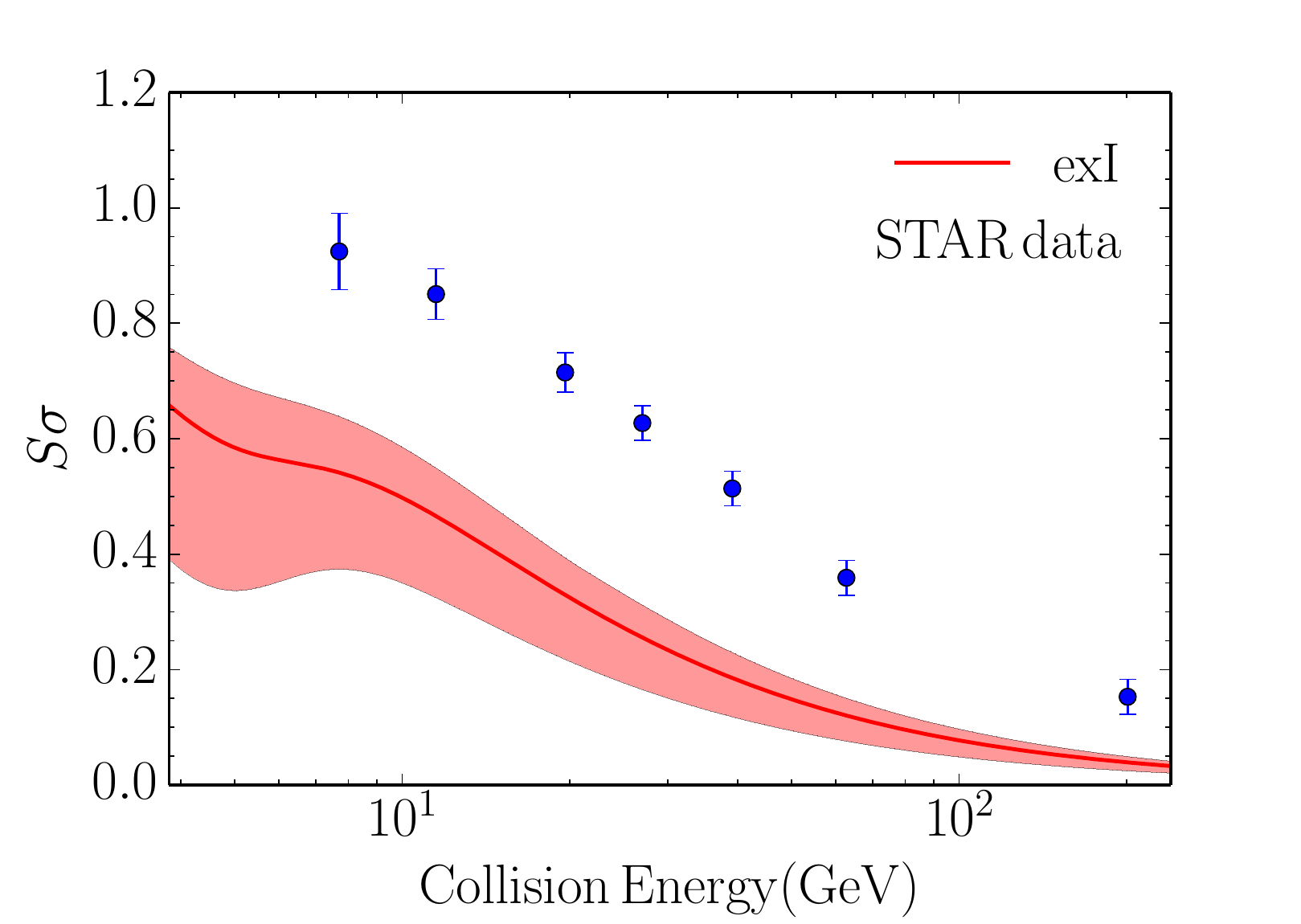}
\includegraphics[width=0.8\linewidth]{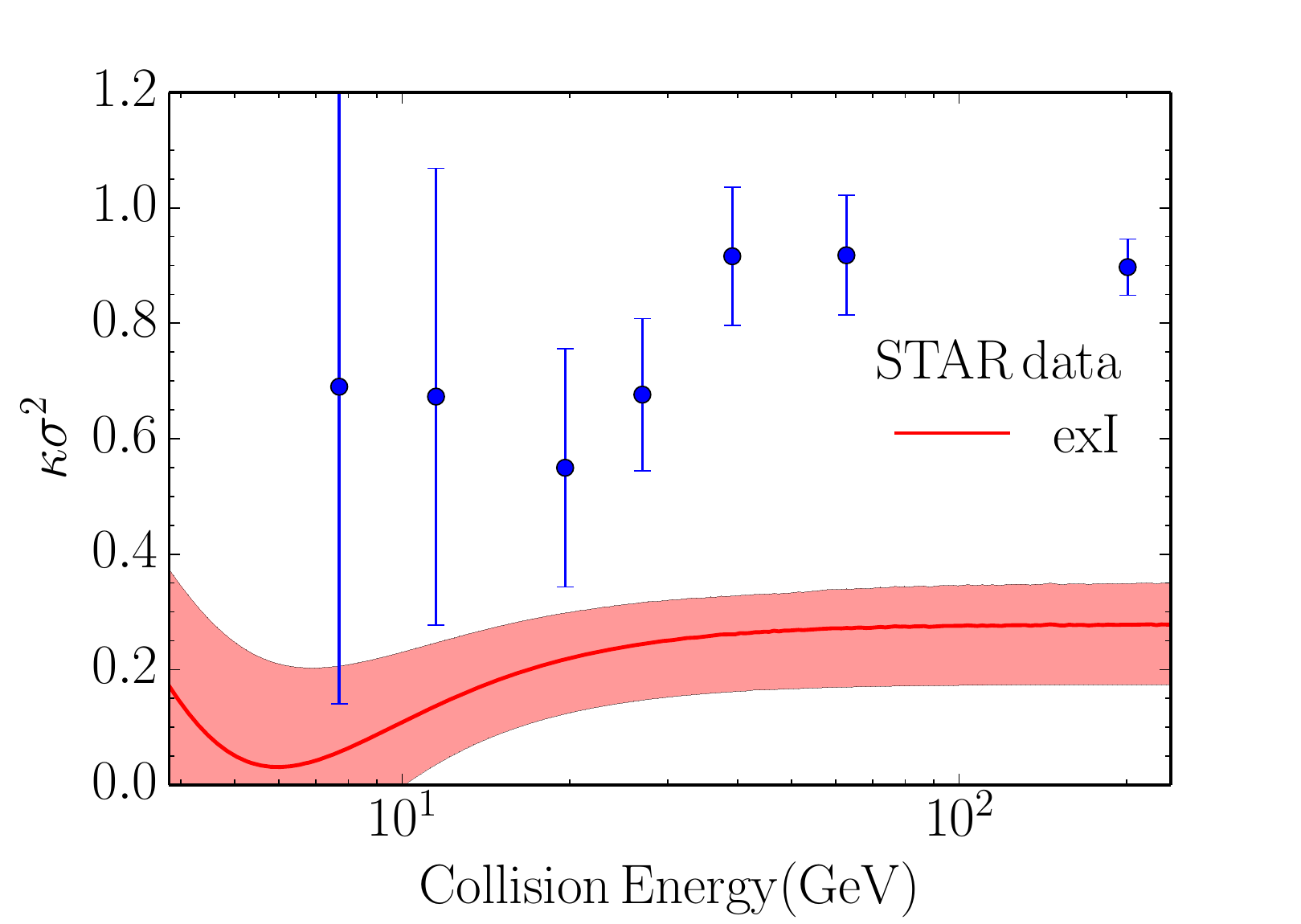}
\caption{(color online) The skewness (top panel) and kurtosis (bottom panel) for the crossover model exI.  The shaded region shows the uncertainty when fitting the crossover equation of state parameters to lattice QCD at zero chemical potential.  The energy dependence of the temperature and chemical potential are determined by the conditions at average chemical freeze-out as in Eq.~(\ref{Tchem}).}
\end{center}
\label{crossover_kurtosis_STAR_166errorband}
\end{figure}

\section{Comparison to STAR Data: Post-Chemical Freeze-Out}
\label{postchemical}

The lack of agreement between the crossover equations of state and the STAR data is significant.  We have already compared the crossover equations of state with lattice QCD in detail previously, with very positive results.  One possibility, which we explore here, is that the baryon fluctuations are not frozen out at the same time as average chemical freeze-out.  Indeed, it is well known that kinetic freeze-out occurs after average chemical freeze-out when the temperature is lower \cite{kinetic1, kinetic2}.  A hint in this direction is that both the skewness and kurtosis decrease with increasing temperature when the chemical potential is fixed.  To get closer agreement with the STAR data, we ought to consider the possibility that the baryon fluctuations freeze-out occurs at lower temperatures than assumed earlier.

As an example, consider changing the parameter $a$ in Eq. \ref{Tchem} from 166 to 140 MeV.  Such a lower temperature is roughly consistent with kinetic freeze-out \cite{kinetic1,kinetic2}.  The results are shown in Fig. 8 for the crossover equation of state exI (the results for exII are nearly identical).  Now the agreement is quite acceptable.  The exceptional points are the kurtoses at the three highest beam energies for which the theory is below the data.  This may very well be associated with the fact that the kurtosis for the excluded volume crossover equations of state, at zero chemical potential, lie below the lattice results, as shown in Figs. 3 and 4.

It is interesting to plot the ratio $\kappa/S^2$ since this is independent of the variance.  The result, using $a = 140$ MeV as in Fig. 8, is shown in Fig. 9.  The agreement at the four lowest beam energies is excellent.  Again, the biggest discrepancies are at the highest beam energies.  As there is no discrepancy for $\sqrt{s_{NN}}$ = 7.7, 11.5, 19.6, and 27 GeV, it would be difficult to argue for a critical point in this energy range - at least under the assumptions made here.

At each beam energy there are two experimental measurements: $\kappa \sigma^2$ and $S \sigma$.  Assuming an equation of state, one can always find a $T$ and $\mu$ at each energy to fit this data.  Using the crossover equation of state exI, we show the results in Fig. 10.  Three features in Fig.~10 are notable.   
The first is that $T$ initially increases with $\mu$. This feature is easily explained from the STAR data.  From Fig. 6, we see that as $\sqrt{s_{NN}}$ decreases below 200 GeV (hence, as $\mu$ increases), the experimentally-measured $\kappa\sigma^2$ decreases. From the lattice results shown in Fig. 3, we know that $\kappa \sigma^2$ is inversely related to $T$ (at least for small $\mu$).  Hence, the inferred $T$ should grow with $\mu$ (for small values of $\mu$).
It is interesting to note that a similar behavior was found in \cite{Alba2014305}, where the point hadron resonance gas model was fit to the variance at all but the lowest energy of $\sqrt{s_{NN}}$ = 7.7 GeV.  The second feature is that $T$ is definitely smaller than inferred from the average chemical freeze-out.  The third is that at large $\mu$, corresponding to small $\sqrt{s_{NN}}$, there is a very large uncertainty in $\mu$.  This uncertainty is a direct consequence of the experimental uncertainties.

\begin{figure}[thp]
\begin{center}
\includegraphics[width=0.8\linewidth]{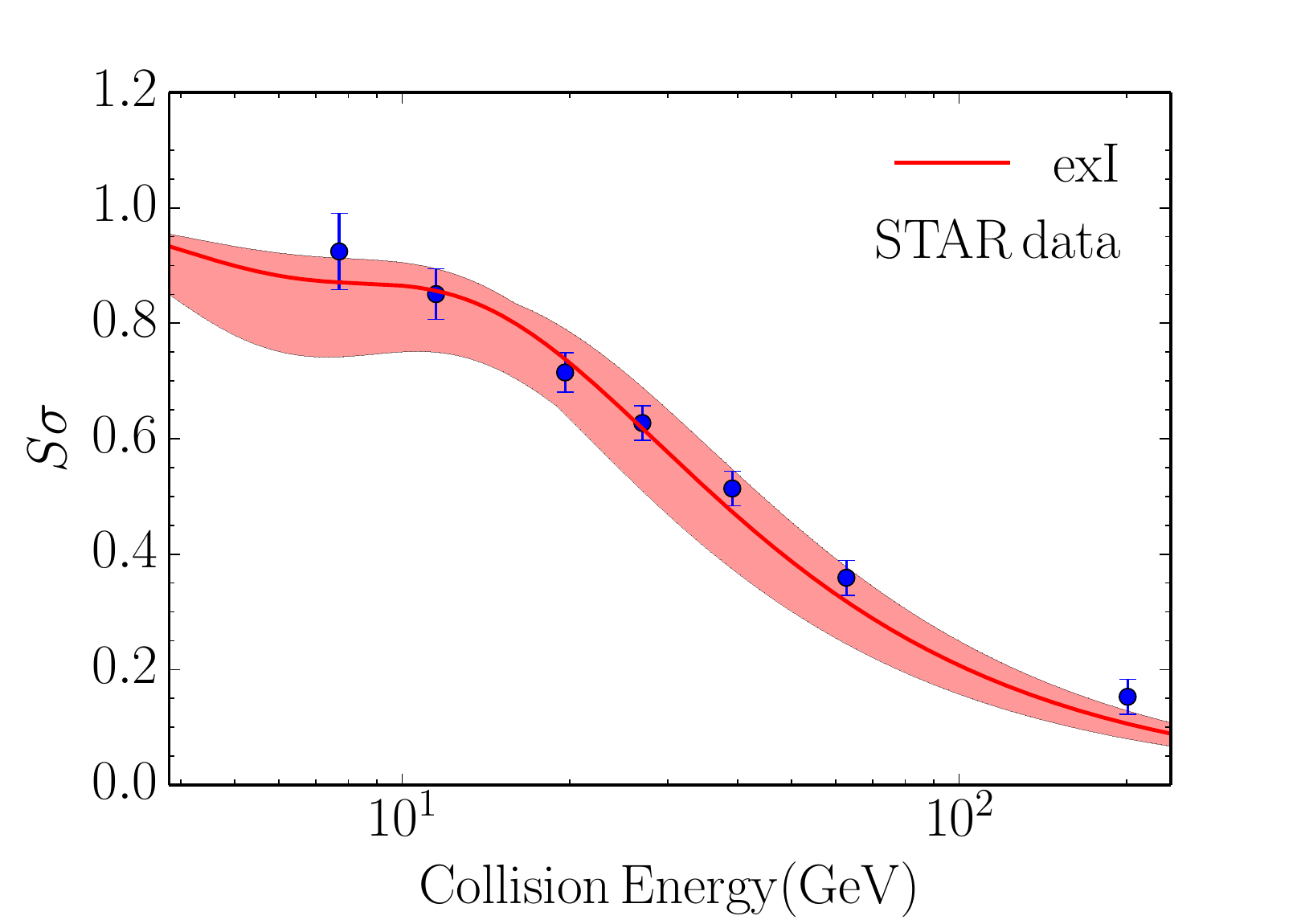}
\includegraphics[width=0.8\linewidth]{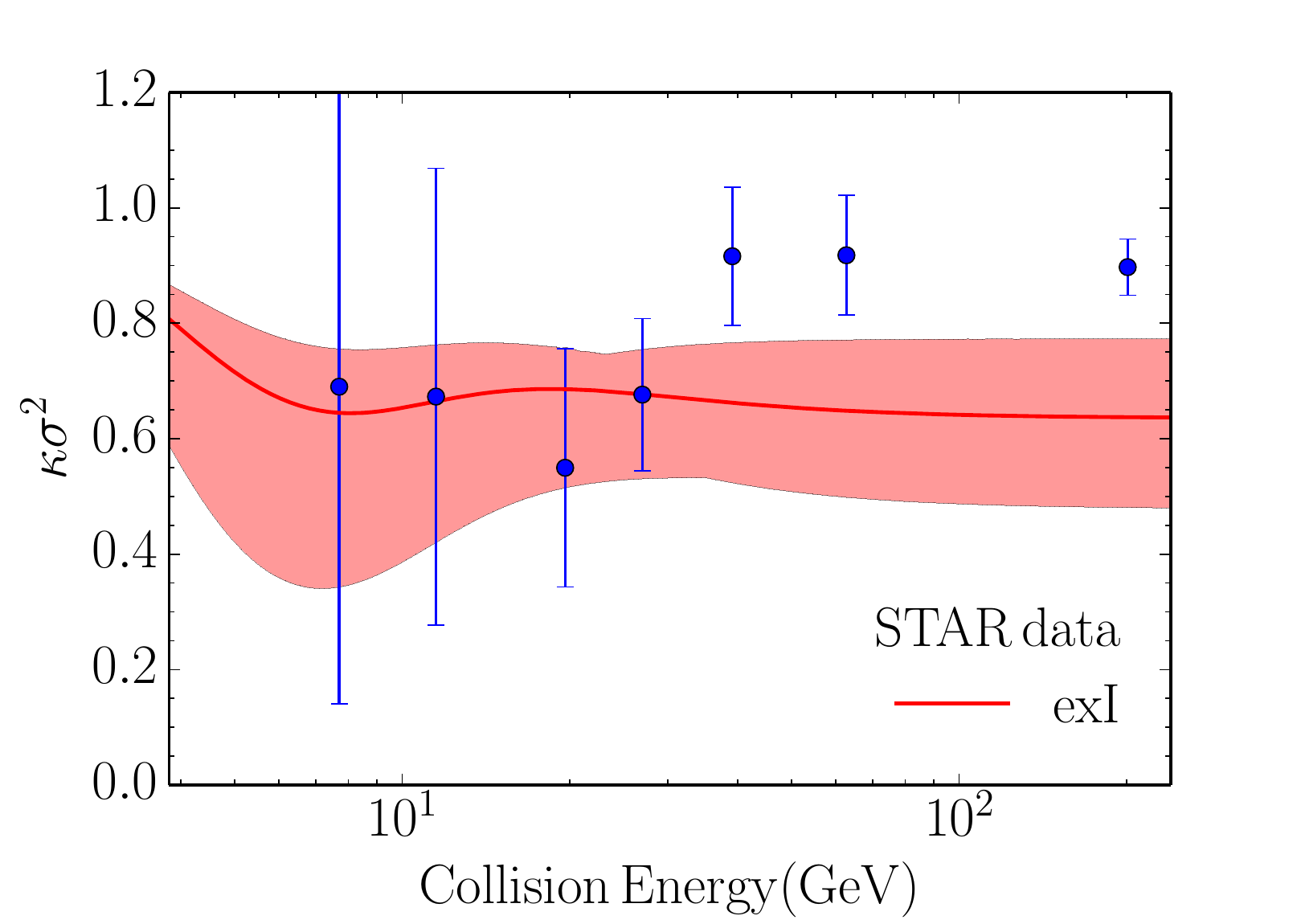}
\caption{(color online) The skewness (top panel) and kurtosis (bottom panel) for the crossover model exI.  The shaded region shows the uncertainty when fitting the crossover equation of state parameters to lattice QCD at zero chemical potential.  The energy dependence of the temperature and chemical potential are determined as in Eq. (\ref{Tchem}) but with a temperature which is 26 MeV lower than in the previous figure ($a = 140$ MeV).}
\end{center}
\label{crossover_kurtosis_STAR_140errorband}
\end{figure}

\begin{figure}[thp]
\begin{center}
\includegraphics[width=0.8\linewidth]{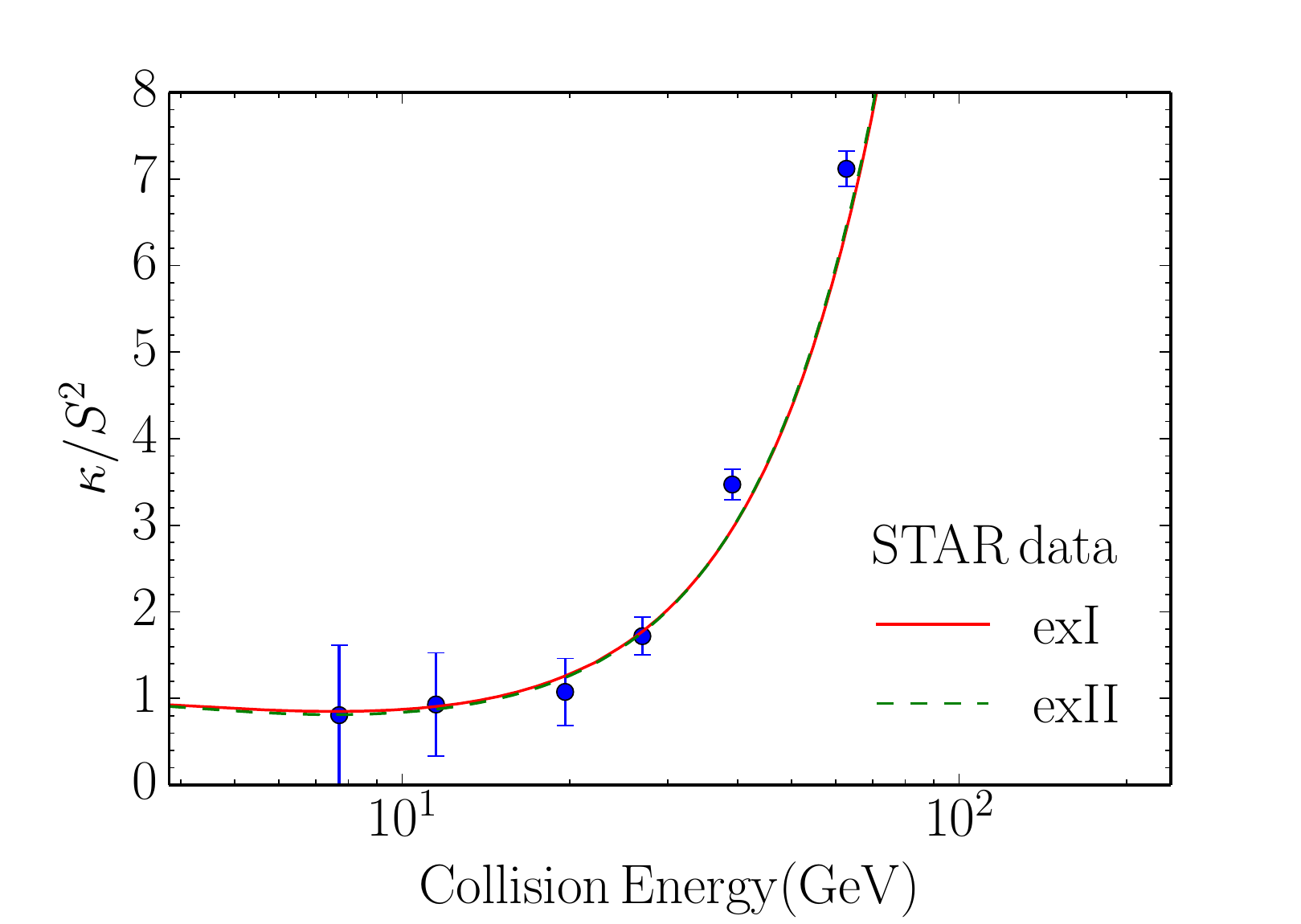}
\caption{(color online) Ratio of kurtosis to the square of skewness for the two crossover equations of state compared to the measurements by the STAR collaboration. }
\end{center}
\label{crossover_ratio_STAR_140}
\end{figure}

\begin{figure}
\begin{center}
\includegraphics[width=0.8\linewidth]{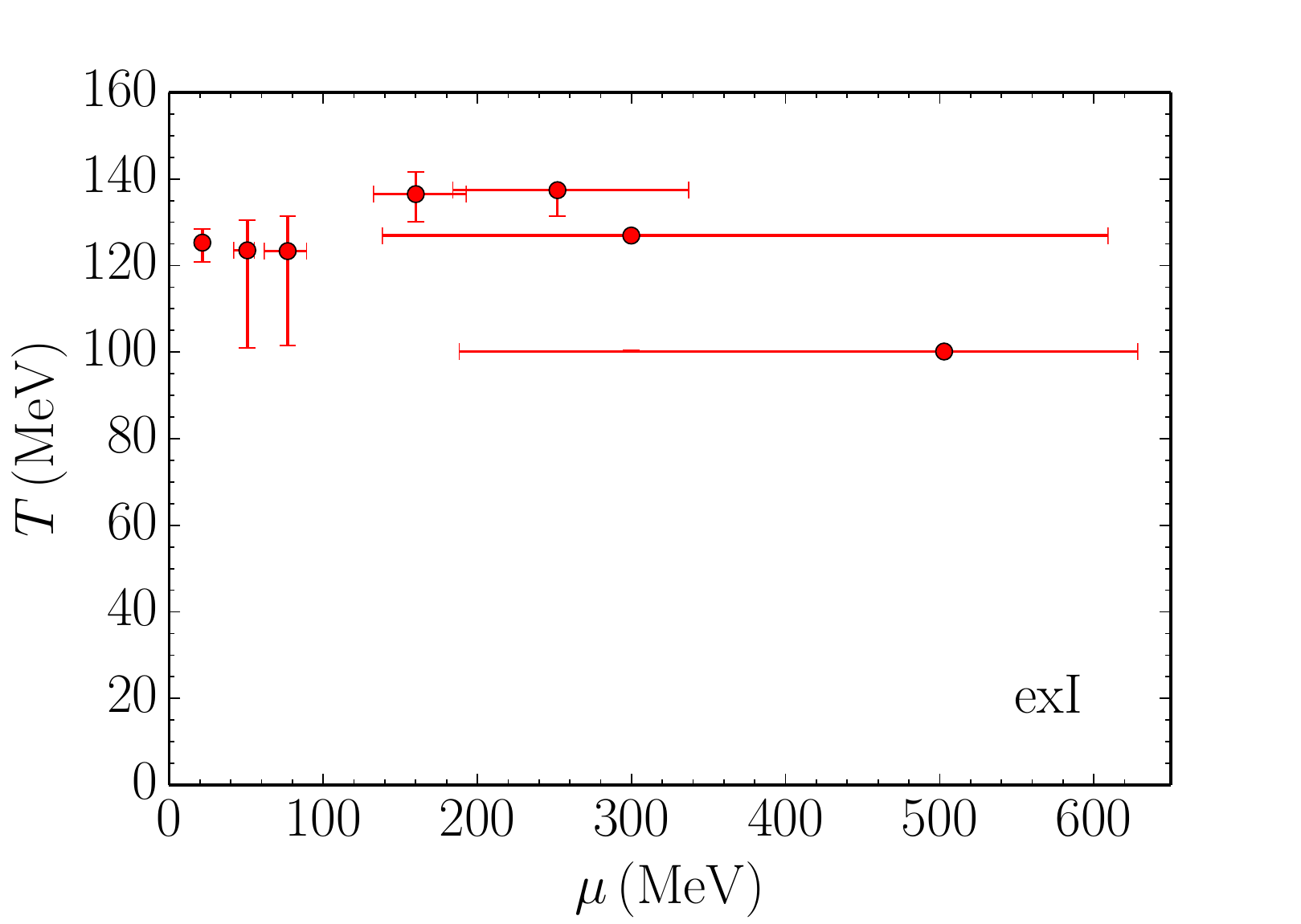}
\caption{(color online) A fit to the STAR measurements of the temperature and chemical potential at each beam energy using the crossover equation of state exI.}
\end{center}
\label{T_vs_mu_STAR}
\end{figure}

\vfill
\newpage

\section{Conclusion}
\label{conclusion}

In this paper, we compared crossover equations of state with lattice QCD results and with the measurements of the first beam energy scan at RHIC.  The previously constructed crossover equations of state interpolated between perturbative QCD at high energy density to a hadronic resonance gas at low energy density.  The hadronic resonance gas, with excluded volume effects included, gave excellent agreement with the sound speed as calculated on the lattice.  Unsatisfactory results were found when hadrons were treated as point particles.

Skewness and kurtosis of the baryon number fluctuations are very sensitive measures of the equation of state because they involve third and fourth derivatives of the pressure with respect to the chemical potential.  The crossover equation of state agreed best with lattice QCD calculations when excluded volume effects in the hadronic sector were taken into account.  Henceforth we rejected the crossover equation of state with point hadrons.

The crossover equations of state are in qualitative agreement but in quantitative disagreement with experimental measurements of the skewness and kurtosis when it is assumed that the fluctuations were frozen out at the same time as average chemical freeze-out.  When baryon fluctuations are allowed to freeze out at lower temperatures, much better agreement is obtained, except for the higher beam energies.  This conclusion is supported by other studies.

There are obvious questions that deserve further investigation.  How accurate are the lattice QCD results, especially at nonzero chemical potential?  How accurately does the crossover equation of state need be known to replicate the lattice QCD equation of state, given that the skewness and kurtosis involve third and fourth order derivatives of the pressure?  Our study does not include the requirement that the system have zero net strangeness, which is probably not a major factor but still needs investigation.  A serious issue is the phenomenology connecting the experimental measurements to the equation of state.  For example, the experimental measurements have a lower momentum-space cutoff for protons of 400 MeV.  Such cutoff effects have been investigated in \cite{Garg2013} and \cite{Bhattacharyya2014}.  However, in general these cutoff effects are not so  straightforward when the equation of state includes interactions.

Our study does not suggest evidence for a critical point, but clearly there is much work to be done, both theoretically and experimentally.

\section*{Acknowledgements}

M. A. and J. K. are supported by the US Department of Energy (DOE) under Grant No. DE-FG02-87ER40328.  C. Y. is supported by the US Department of Energy (DOE) under Grant No. DE-FG02-03ER41259.


\newpage

\section*{References}

\begin{thebibliography}{99}

\bibitem{Stephanov}
M. Stephanov, Prog. Theor. Phys. Suppl. {\bf 153}, 139 (2004); Int. J. Mod. Phys. A {\bf 20}, 4387 (2005); PoS(LAT2006)024.

\bibitem{MohantyQM}
B. Mohanty, Nucl. Phys. A {\bf 830}, 899c (2009).

\bibitem{STAR_BES} L. Adamczyk, {\it et al.} [STAR Collaboration], Phys. Rev. Lett. {\bf 112}, 032302 (2014).

\bibitem{nongaussian} M. A. Stephanov, Phys. Rev. Lett. {\bf 102}, 032301 (2009).

\bibitem{3rdmoments} M. Asakawa, S. Ejiri, and M. Kitazawa, Phys. Rev. Lett. {\bf 103}, 262301 (2009).

\bibitem{4thmoments} F. Karsch and K. Redlich, Phys. Lett. B {\bf 695}, 136 (2011).

\bibitem{matchingpaper} M. Albright, J. Kapusta and C. Young, Phys. Rev. C {\bf 90}, 024915 (2014).

\bibitem{Fukushima} K. Fukushima, Phys. Rev. C {\bf  91}, 044910 (2015) .

\bibitem{Borsanyi2014PRL} S. Bors{\'a}nyi, Z. Fodor, S. D. Katz, S. Krieg, C. Ratti, and K. K. Szab{\'o}, Phys. Rev. Lett. {\bf 113}, 052301 (2014).

\bibitem{pdg2012} J. Beringer {\it et al.} (Particle Data Group), Phys. Rev. D {\bf 86}, 010001 (2012).

\bibitem{Borsanyi2010JHEP} S. Bors{\'a}nyi, G. Endr{\H o}di, Z. Fodor, A. Jakov{\'a}c, S. D. Katz, S. Krieg, C. Ratti, and K. K. Szab{\'o}, J. High Energy Phys. {\bf 11}, 077 (2010).

\bibitem{Borsanyi2012} S. Bors{\'a}nyi, G. Endr{\H o}di, Z. Fodor, S. D. Katz, S. Krieg, C. Ratti, and K. K. Szab{\'o},  J. High Energy Phys. {\bf 08}, 053 (2012).  

\bibitem{Borsanyi2013PRL} S. Bors{\'a}nyi, Z. Fodor, S. D. Katz, S. Krieg, C. Ratti, and K. K. Szab{\'o}, Phys. Rev. Lett. {\bf 111}, 062005 (2013).

\bibitem{freezeoutcurve} J. Cleymans, H. Oeschler, K. Redlich, and S. Wheaton, Phys. Rev. C {\bf 73}, 034905 (2006).

\bibitem{kinetic1} B. Schenke, S. Jeon, and C. Gale, Phys. Rev. C {\bf 82}, 014903 (2010).

\bibitem{kinetic2} C. Shen, U. W. Heinz, P. Huovinen, and H. Song,  Phys. Rev.C {\bf 82}, 054904 (2010); {\it ibid.} {\bf 84}, 044903 (2011).

\bibitem{Alba2014305} P. Alba, W. Alberico, R. Bellwied, M. Bluhm, V. Mantovani Sarti, M. Nahrgang, and C. Ratti, Phys. Lett. B {\bf 738}, 305 (2014).

\bibitem{Garg2013} P. Garg, D. K. Mishra, P. K. Netrakanti, B. Mohanty, A. K. Mohanty, B. K. Singh, and N. Xu, Phys. Lett. B {\bf 726}, 691 (2013).

\bibitem{Bhattacharyya2014} A. Bhattacharyya, S. Das, S. K. Ghosh, R. Ray, S. Samanta, Phys. Rev. C {\bf 90}, 034909 (2014).



\end{thebibliography}
\end{document}